\documentclass{eas}
\usepackage{graphicx}
\begin{document}

\title{Groups of dwarf galaxies in the Local supercluster} 
\author{Dmitry Makarov}
\address{Special Astrophysical Observatory of the Russian Academy of Sciences, Russia; \email{dim@sao.ru}}
\secondaddress{Universit\'e de Lyon, Universit\'e Lyon 1, CNRS/IN2P3, Institut de Physique Nucl\'eaire de Lyon, Villeurbanne, France}
\author{Igor Karachentsev}
\sameaddress{1}
\author{Roman Uklein}
\sameaddress{1}
\runningtitle{Makarov \etal: Groups of dwarfs}
\begin{abstract}
We present a project on study of groups composed of dwarf galaxies only. 
We selected such structures using HyperLEDA and NED databases 
with visual inspection on SDSS images and on digital copy of POSS. 
The groups are characterized by size of few tens of kpc and 
line-of-sight velocity dispersion about 18 km s$^{-1}$. 
Our groups similar to associations of nearby dwarfs from Tully et al.\ (2006). 
This specific population of multiple dwarf galaxies such as I~Zw~18 
may contain significant amount of dark matter. 
It is very likely that we see them at the stage just before merging of its components. 
\end{abstract}
\maketitle

\section{Introduction}
\label{intro}

The modern mass surveys of galaxy redshifts like 
2dF (Colless \etal\ \cite{2dF}), HIPASS (Zwaan \etal\ \cite{HIPASS}), 
6dF (Jones \etal\ \cite{6dF}), ALFALFA (Giovanelli \etal\ \cite{ALFALFA}) and 
SDSS (Abazajian \etal\ \cite{SDSS})
give us extensive opportunities to refine and improve our knowledge about the structure of our Universe.
In the series of papers (Karachentsev \& Makarov \cite{Pairs}; Makarov \& Karachentsev \cite{Triplets,Groups}) 
we have studied the distribution and properties of galaxy groups on scale of the Local supercluster. 
During this work we found surprisingly high 
fraction of groups consisting of dwarf galaxies.
Some very interesting objects happen to be among these groups. 
It is famous metal deficient galaxy I~Zw~18. 
The pair of extremely metal-poor blue compact dwarf HS~0822+3542 and low surface brightness object SAO~0822+3545 
were studied by Chengalur \etal\ (\cite{HS0822+3542}).
It seems that gas rich galaxies with very low metallicity (Ekta \etal\ \cite{SBS1129+576}) appear
quite often among the systems from our sample.
On the other hand, Tully \etal\ (\cite{TullyAssoc}) pointed out the existence in the Local Volume
of associations of galaxies which exclusively consist of dwarfs.
That associations were identified based on 3D map of nearby galaxies with distances of high precision.
These structures can contain big amount of dark matter and 
the mass-to-light ratios are in the range 100--1000 M/L in solar units.
Our systems are related to Tully's association of dwarfs and they also could have very high mass-to-light ratio.

We have started spectroscopic survey of galaxies from our sample on Russian 6-meter telescope of SAO RAS.
In this work we describe the selection and analyse the properties of groups of dwarf galaxies in the Local supercluster.

\section{The data}
\label{sec:data}

We use the HyperLEDA\footnote{http://leda.univ-lyon1.fr} 
(Paturel \etal\ \cite{HyperLEDA}) and 
the NED\footnote{http://nedwww.ipac.caltech.edu} databases
as main sources of data on radial velocities, apparent
magnitudes, morphological types and other parameters of galaxies.
A blind use of database is fraught with false and erroneous data.
Both databases contain a significant amount of `spam': 
objects with erroneous radial velocities that come from the 
mass sky surveys such as 2dF etc. 
Quite common case is a confusion of coordinates and velocities 
of galaxies located close to each other on the sky. 
Apparent magnitudes and radial velocities from the SDSS
survey often correspond to individual knots and associations in
bright galaxies. 
We have taken into account and corrected different kind of contamination of the databases.
As a matter of fact it is most hard and time-consuming part of our work.

Additionally, we made a number of optical identifications of
$HI$ sources from the HIPASS survey, specifying their coordinates
and determining the apparent magnitudes and morphological types of
galaxies (Karachentsev \etal\ \cite{KMKM2008}). 
Many dwarf galaxies, especially of low surface brightness, 
were examined by us on the DSS digital images to determine 
their magnitudes and morphological types. 

We use $K$-band photometry as indicator of stellar mass of a galaxy because
it is weakly affected by a dust and young blue star complexes in the galaxy. 
Most of near-infrared photometry comes from the 
all-sky 2MASS survey (Jarrett \etal\ \cite{2MASSX,2MASSAtlas}).
In case of lack of $K$-band photometry 
we transferred the optical ($B,V,R,I$) and near infrared ($J,H$) 
magnitudes into the $K$-magnitudes
as it described in series of our papers (Karachentsev \& Makarov \cite{Pairs}; Makarov \& Karachentsev \cite{Triplets,Groups}).
Note that because of short exposure the 2MASS survey turned out to
be insensitive to the galaxies with low surface brightness and blue colour. 
Thus we have the direct near-infrared measurements for about 65\% of galaxies in our sample and
for 35\% the $K$ magnitude was estimated from optic.

We collected 10914 galaxies with radial velocities in the Local Group
rest frame of $V_{LG}<3500$ km s$^{-1}$, located at the galactic
latitudes $|b|>15^{\circ}$. 
The sample of such a depth contains the entire
the Local supercluster with its distant outskirts, surrounding voids
and ridges of the  neighbouring clusters.

\section{The algorithm}
\label{sec:algorithm}

Our algorithm (Karachentsev \cite{K1994}; Makarov \& Karachentsev \cite{MK2000})
for group selection is based on natural requirement
that total energy of physical pair of galaxies has to be negative.
\begin{equation}
\frac{V_{12}^2R_{12}}{2GM_{12}}<1,
\label{eq:Criterion1}
\end{equation}
where $M_{12}$ is the total mass of the pair, and $G$ is the
gravitational constant. However, observations give us only
radial velocities and the sky-plane projected distance between the galaxies.
Two galaxies with a very small difference in radial
velocities but a large separation in the sky can meet the
condition (\ref{eq:Criterion1}) without being mutually bound. 
Hence the condition of negative total energy of the pair, 
expressed in terms of the observables
must be added by another restriction on the maximal
distance between the components at their fixed mass $M_{12}$. 
The condition when the pair components remain within the sphere of
`zero-velocity' (Sandage \cite{Sandage1986}) takes the form of
\begin{equation}
 \frac{\pi H_0^2R^3_{\bot}}{8GM_{12}}<1,
\label{eq:Criterion2}
\end{equation}
where $H_0$ is the Hubble constant.

We determined the masses of galaxies from their integral
luminosity in the infrared $K_s$-band, supposing that they have
the same mass-to-luminosity ratio
\begin{equation}
  M/L_K={\kappa} (M_{\odot}/L_{\odot}),
\label{eq:ML}
\end{equation}
where $\kappa$ is taken equal to 6. 
In the fact, the value of $\kappa=6$ is only more or less 
arbitrary parameter of the algorithm. 
To bound it we `trained' the clusterization algorithm 
(\ref{eq:Criterion1}--\ref{eq:ML}) on detailed three-dimensional
distribution of galaxies in the Local Volume (Karachentsev \etal\ \cite{CNG}), 
where the membership of galaxies in the groups is known from good quality
photometric distances. 
The choice of $\kappa=6$ is the compromise between a loss of the real members
and an impurity of groups by false members.
For the $\kappa\le4$ we lose significant number of real members
while $\kappa\ge8$ leads to appearance in the groups suspicious members.
Moreover, for $\kappa\ge10$ galaxies are combined into extended 
non-virialized aggregates. 
At the given value of $\kappa=6$ the
dwarf companions in the well-known nearby groups are usually
located inside the zero velocity surface around the major galaxies
of these groups.

\section{Basic properties of the groups of dwarfs}

\begin{figure}
\begin{tabular}{cc}
\includegraphics[width=0.47\textwidth]{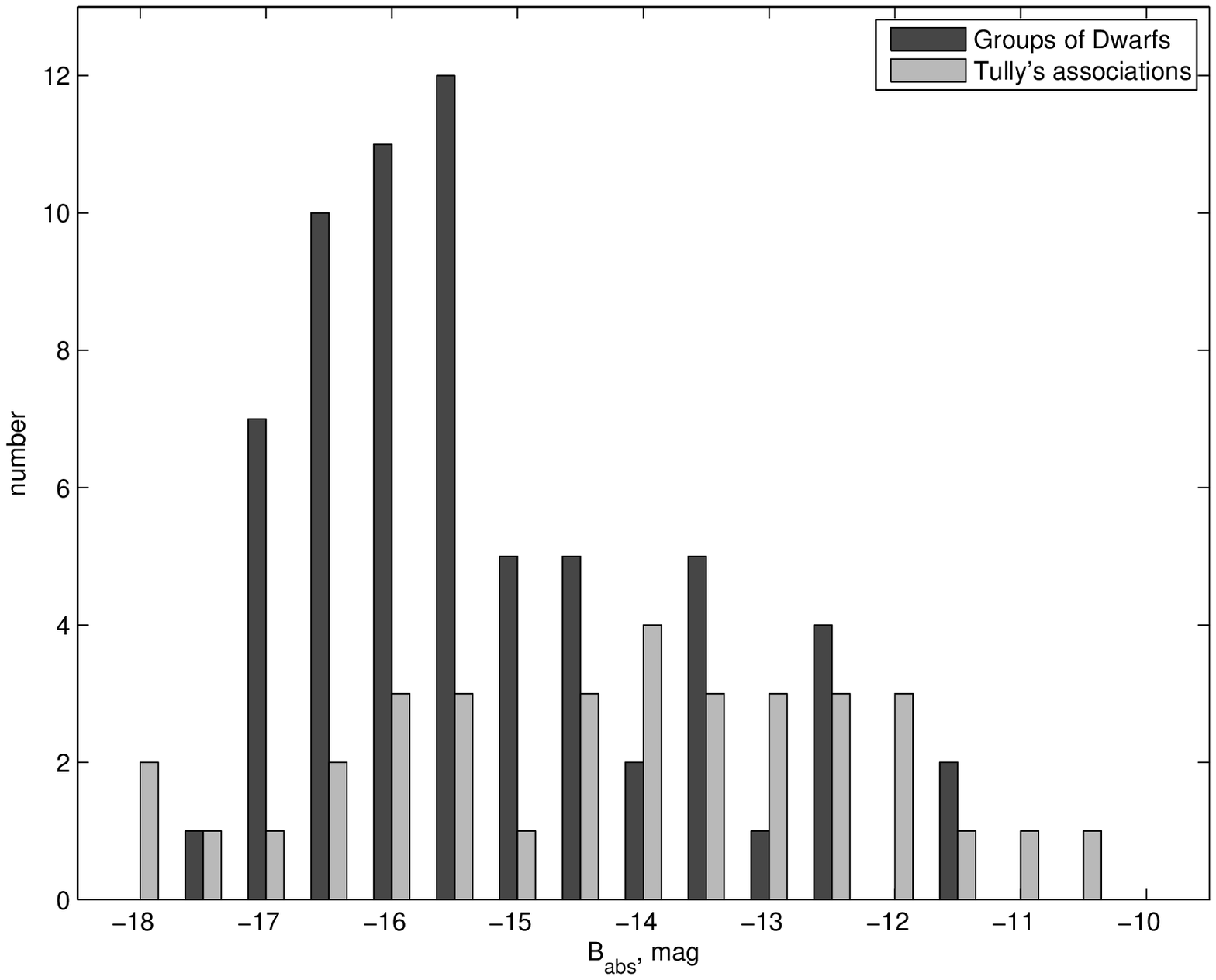}
&
\includegraphics[width=0.47\textwidth]{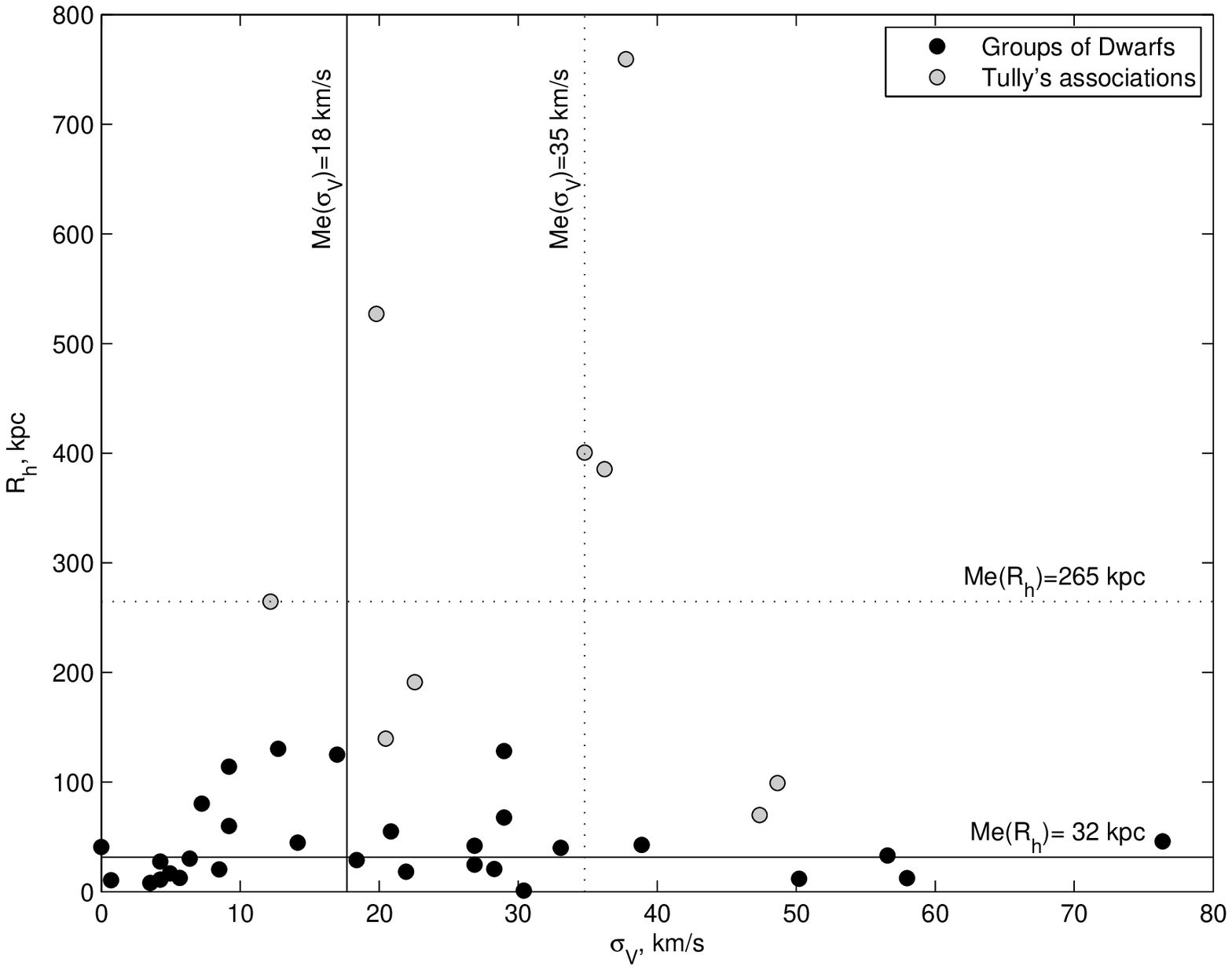}
\end{tabular}
\caption{
The left panel shows luminosity function of galaxies in comparison 
with associations of dwarfs (Tully \etal\ \cite{TullyAssoc}).
The distribution of velocity dispersion and harmonic radius 
of the groups of dwarfs in comparison with Tully's associations 
is shown on the right panel.}
\label{fig:h_babs}
\label{fig:rh_sv}
\end{figure}

This section contains short description of main properties of groups in the Local supercluster 
(Makarov \& Karachentsev \cite{Groups}).
We gathered 5926 objects of 10914 galaxies in 1082 groups.
The 395 groups have a population of $n\geq4$ members.
The dispersion ($\sigma_V$) in well populated ($n\geq4$) groups ranges 
from 10 to 450 km s$^{-1}$ with the median of 74 km s$^{-1}$. 
The mean harmonic radius of the groups is distributed over 
a wide range from 7 kpc to 750 kpc with a median of 204 kpc.
The median crossing time of selected systems is 2.2 Gyr. 
Only 2\% of groups fall have estimation of crossing time higher than age of Universe (13.7 Gyr). 
Consequently, almost all of the selected by our criterion groups can be considered as dynamically
evolved systems.
The groups are characterized by median mass corrected for measurement error $M_p^c=2.3\,10^{12}M_{\odot}$
and mass-to-luminosity ratio in $K$-band $M/L=22$ $M_{\odot}/L_{\odot}$.

The special interest are drawn by the groups where brightest member has 
absolute magnitude below $M_K=-19$ (the luminosity of the SMC).
Surprisingly, the sample of groups of dwarf galaxies contains 30 such systems.
It corresponds to at least 3\% of the all groups in the Local supercluster.
The distribution of the groups over the sky is highly inhomogeneous.
We have found only 2 systems outside of zone of SDSS survey.
Our sample of galaxies is subjected to different kind of observation selection.
Therefore, it is impossible to estimate the incompleteness of the sample, 
but it seems that the number of such systems should be quite significant.
The dispersion of radial velocities in groups of dwarfs ($\sigma_V$) 
is less than 80 km s$^{-1}$ with the median of 18 km s$^{-1}$. 
The projected size of groups is less than 200 kpc.
The median values of harmonic radius is 32 kpc.
Thus, the groups of dwarfs are much more compact and 
they have significantly smaller velocity dispersion 
than normal groups in the Local supercluster.

\begin{figure}
\begin{tabular}{cc}
\includegraphics[width=0.47\textwidth]{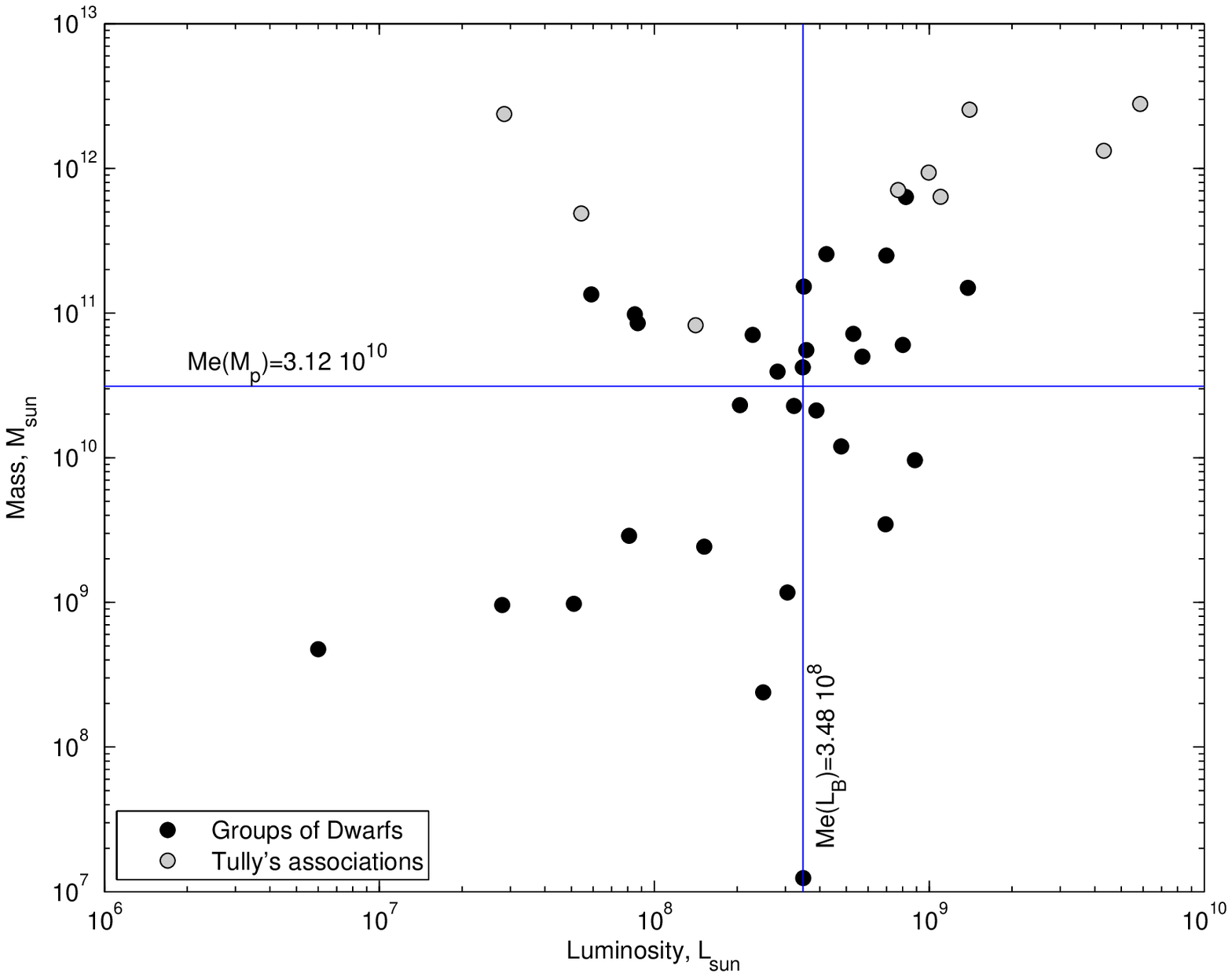}
&
\includegraphics[width=0.47\textwidth]{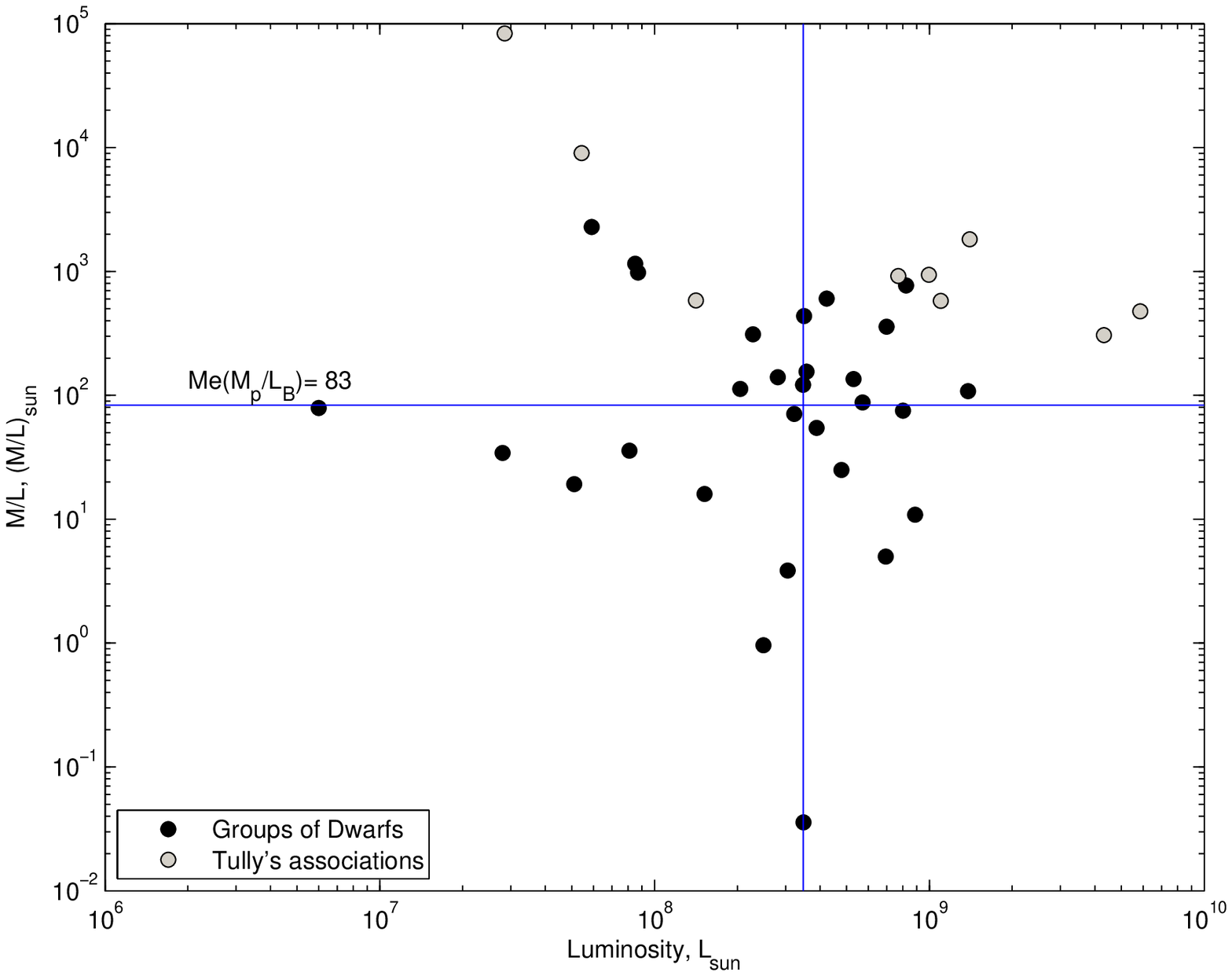}
\end{tabular}
\caption{
The mass-luminosity relation for groups under consideration is presented on left panel.
The mass-to-light ration versus luminosity for sample of groups of dwarf galaxies is shown on right panel.}
\label{fig:mass_lum}
\label{fig:ml_lum}
\end{figure}

The luminosity function of the groups of dwarfs is shown on Fig.~\ref{fig:h_babs} (left panel).
The groups of dwarfs (dark gray bars) occupy the same range of absolute magnitudes 
as the Tully's associations of dwarf galaxies (light gray bars) (Tully \etal\ \cite{TullyAssoc}). 
Unlike to nearby galaxies our sample is very incomplete.
It explains the sharp drop of the number of galaxies bellow $M_B=-16$ absolute magnitude.
The Fig.~\ref{fig:rh_sv} (right panel) illustrates the relation between sizes and 
velocity dispersions in the groups and associations.
The groups and associations have quite comparable velocity dispersion 18 and 35 km s$^{-1}$ respectively,
while the size of groups (32 kpc) is significantly smaller than size of associations (265 kpc).
This big difference in sizes of groups and associations is explained by different method of system selection.
Our algorithm is oriented to find a bounded and virialized groups of galaxies, while the associations of dwarf were selected 
by correlation in position, velocity and distances of nearby galaxies.
The median value of luminosity of the groups is $3.5\,10^8$ $L_\odot$ in $B$-band and median of mass is $3.1\,10^{10}$ $M_\odot$.
It lead to mass-to-light ratio of 83 in solar units.
Despite of mass of the groups is systematically lower then mass of the association,
the groups of dwarfs form continuous sequence with Tully's association (see Fig.~\ref{fig:mass_lum}).

\section{Conclusion}

In the last decade the modern mass redshift survey significantly increase 
the number of galaxies with known velocity in the Local supercluster.
We developed the algorithm of group selection which is base on 
on natural requirement that total energy of bounded pair of galaxies has to be negative 
(Karachentsev \& Makarov \cite{Pairs}; Makarov \& Karachentsev \cite{Triplets,Groups}).
Application of our method to new data has allowed us to find interesting kind of groups 
which consist of only dwarf galaxies. The number of such groups is surprisingly high.
They amount at least 3\% of the all groups in the Local supercluster.

Most of the galaxies in our sample are blue and show sign of ongoing star formation.
The galaxies of very low metallicity like I~Zw~18 or HS~0822+3542 appear quite often among of them
(Chengalur \etal\ \cite{HS0822+3542}; Ekta \etal\ \cite{SBS1129+576}).

The groups of dwarfs are characterized by size of 32 kpc and velocity dispersion of 18 km s$^{-1}$.
These values are significantly smaller then sizes and inner motion in ordinary groups in the Local supercluster 
(204 kpc and 74 km s$^{-1}$, respectively). The groups of dwarfs form continuous sequence with association of dwarfs
which were found by Tully \etal\ (\cite{TullyAssoc}) using precise distance determination. 
The groups have about the same luminosity and velocity dispersion as association but associations are significantly wider.
The median value of mass-to-light radio of groups of dwarfs is 83 $M_\odot/L_\odot$. 
It indicates that such kind of groups contains significant amount of dark matter.

\begin{acknowledgements}
This work was supported by Russian Foundation
for Basic Research grants 08--02--00627 and 08--02--90402. 
\end{acknowledgements}


\end{document}